\RequirePackage{ifpdf}
\ifpdf 
\documentclass[pdftex]{sigma}
\else
\documentclass{sigma}
\fi

\def\pd#1#2{\frac{\partial#1}{\partial #2}}
\def\ppdd#1#2#3{\frac{\partial^2#1}{\partial #2\partial#3}}
\def\pppddd#1#2#3#4{\frac{\partial^3#1}{\partial #2\partial#3 \partial #4}}
\def\lR{\mathbb{R}}
\def\cD{{\cal D}}
\def\cE{{\cal E}}
\def\cX{{\cal X}}

\begin{document}

\renewcommand{\PaperNumber}{032}

\FirstPageHeading

\renewcommand{\thefootnote}{$\star$}

\ShortArticleName{Equivariance, Variational Principles, and the Feynman Integral}

\ArticleName{Equivariance, Variational Principles,\\ and the Feynman Integral\footnote{This paper is a contribution to the Proceedings
of the Seventh International Conference ``Symmetry in Nonlinear
Mathematical Physics'' (June 24--30, 2007, Kyiv, Ukraine). The
full collection is available at
\href{http://www.emis.de/journals/SIGMA/symmetry2007.html}{http://www.emis.de/journals/SIGMA/symmetry2007.html}}}

\Author{George SVETLICHNY}

\AuthorNameForHeading{G. Svetlichny}

\Address{Departamento de Matem\'atica, Pontif\'{\i}cia Unversidade Cat\'olica, Rio de Janeiro, Brazil}

\Email{\href{mailto:svetlich@mat.puc-rio.br}{svetlich@mat.puc-rio.br}}

\URLaddress{\url{http://www.mat.puc-rio.br/~svetlich/}}

\ArticleDates{Received November 02, 2007, in f\/inal form March
13, 2008; Published online March 19, 2008}

\Abstract{We argue  that the variational calculus leading to Euler's equations and Noether's theorem can be replaced by equivariance and invariance conditions avoiding the action integral. We also speculate about the origin of Lagrangian theories in physics and their connection to Feynman's integral.}


\Classification{49N99; 49Q99; 58D30; 58K70; 70S05; 70S10}

\section{Introduction}

This paper is part of an on-going investigation into Lagrangian theories in an attempt to understand why they are so prevalent in physics. Part of the mystery is that though one uses variation of the action integral to get the equations of motion or conservation laws (via Noether's theorem), the actual convergence of the integral is generally not relevant, unless one is {\em really\/} trying to extremize the value, such as f\/inding the shortest path or smallest area. Such manipulations are formal yet very useful. When one comes to quantization, in the Feynman integral approach the integrand is a phase with the action integral in the exponent. The Feynman integral again is used as a formal object whose very def\/inition is unclear. Why are these procedures so successful?

We won't be able to answer this question, only hope to give some hints of an answer. A~convenient starting hypothesis  which would explain the ubiquity of Lagrangian theories is to consider all successful Lagrangian theories as ef\/fective theories arising from some truly fundamental theo\-ry by relegating some degrees of freedom to the background.   The fundamental theory is such that its successful ef\/fective theories are of Lagrangian type. The success of Lagrangian theories is thus due to a specif\/ic nature of the fundamental theo\-ry. Lagrangian theories have the f\/lexibility of hiding and revealing degrees of freedom thus marking a necessary property of any fundamental theory: it must be possible to extract ef\/fective theories from it. It must be possible to ef\/fectively deal only with small and well chosen combinations of variables in the whole set of degrees of freedom that make up the world.

This is also part of an attempt to replace the calculus of variation by geometric constructs and reinterpret the Feynman integral as something other than integration. The hope is that this may shed light on the structure of fundamental physical theories and what makes the successful ones succeed.

In Section~\ref{sec:geom} we present the geometric structure of the calculus of variation in bundle-theoretic terms. This is fairly standard and well known material, much of which can be found in Saunders~\cite{saun}. See also Olver~\cite{olv} for related topics. In Section~\ref{sec:equiv} we re-examine Euler's equations from the point of view of equivariance of certain bundle maps, deducing, as a new result, what all such are. Section~\ref{sec:noeth} addresses Noether's theorem under a new perspective, that by a ``de-ef\/fectivation'', that is, the introduction of equivalent Lagrangians with extra degrees of freedom, variational invariance can be re-expressed as ordinary dif\/feomorphism-induced invariance of the  Lagrangian function. This frees Noether's results from a reliance on the action integral. Section~\ref{sec:conc}, the last, of\/fers some remarks on the whole program and its relation to the Feynman integral. Further discussion of the Feynman integral is to be found in a separate article~\cite{svetfmubs}.

\section{Geometry of the variational calculus}\label{sec:geom}

Let $M$ be a dif\/ferentiable manifold, which we shall  take to be the conf\/iguration space of some classical physical system with a f\/inite number of degrees of freedom. For simplicity's sake we shall not deal with f\/ield theories,  the arguments here should be extensible to the f\/ield theoretic case also. The cotangent bundle $T^*M$ is then what is known as {\em phase space\/}. We shall designate a generic point of phase space by $(q,p)$. The tangent bundle $TM$ is the space of ``positions and velocities'' whose generic point we shall designate by $(q,v)$ or $(q,\dot q)$. We shall deal with the iterated bundles $T^2M=T(TM)$, $T^*(TM)$, $T(T^*M)$, and $T^{*2}M=T^*(T^*M)$.

It is useful to see how various objects look in bundle trivializations,
in particular those induced by a choice of local coordinates
$q^1,\dots,q^n$ in an open set $U\subset M$. In relation
to such local coordinates a typical vector and co-vector in coordinate basis are
\[
v=\sum_iv^i \frac{\partial}{\partial q^i}, \qquad \alpha = \sum_jp_jdq^j.
\]
In what follows we shall adopt a summation rule by which repeated indices, one lower and one upper, are to be summed over from $1$ to $n$ where $n$ is the dimension of the manifold $M$. Note that the index $i$ in the expression $\partial/\partial q^i$ is to be considered as lower.

In the four doubly iterated bundles, using coordinate bases again,
generic points will be denoted as follows
\begin{center}
\renewcommand\arraystretch{1.5}
\begin{tabular}{c|c|c}
Bundle & Generic Point & Abbreviation \\ \hline
\tsep{2ex}\bsep{2ex}$T(TM)$&$  \left(q^i, v^j, U^a\frac{\partial}{\partial q^a}+
V^b\frac{\partial}{\partial v^b}\right)$& $(q, v; U, V)$\\
\bsep{2ex} $T^*(TM)$&$  \left(q^i, v^j, A_a\, dq^a +
B_b\, dv^b\right)$& $(q, v; A,B)$\\
\bsep{2ex} $T(T^*M)$&$  \left(q^i, p_j, S^a\frac{\partial}{\partial q^a}  +
T_b\frac{\partial}{\partial p_b}\right)$&$(q, p; S,T)$ \\
 $T^*(T^*M)$&$  \left(q^i, p_j, Q_a\, dq^a  +
R^b\, dp_b\right)$&$(q, p; Q,R)$
\end{tabular}
\renewcommand\arraystretch{1}
\end{center}

One has to be careful in dealing with these expressions. The eight quantities $U^i$, $V^j$, $A_i$, $B_j$, $S^i$, $T_j$, $Q_i$, and $R^j$ don't necessarily transform under coordinate change in the way the indices suggest. $U$ is a vector and $B$ a co-vector, but in general the transformation properties are more complicated, a fact that will be important.

There is a rather remarkable canonical isomorphisms
{\em as bundles over $M$\/}
\[
\Lambda:T^*(TM)\to T(T^*M ),
\]
which  plays a central role in the variational calculus.
In local coordinates this is
\[
\Lambda: (q,v; A,B) \mapsto (q,B; v,A).
\]

To check that this is coordinate independent is a tedious and unenlightening exercise. There seems to be no way to def\/ine it without recourse to local coordinates and we suspect that it may in fact be impossible to def\/ine it any other way.

There are two projections $T(TM)\to TM$ given by $\pi:(q,v;U,V)\mapsto (q,v)$ and $\pi_*:(q,v;U,V)\mapsto (q,U)$. There is a subbundle $E(TM)\subset T(TM)$ of elements satisfying $\pi(X)=\pi_*(X)$, that is $U=v$. Sections $X$ of $E(TM)$ over $M$ are vector f\/ields whose f\/low is the equivalent f\/irst order system of {\em  second order\/} ordinary dif\/ferential equations. The f\/low def\/ined by $(v,v,V)$ would be $\dot q=v,\,\dot v=V$, that is $\ddot q =V$. So if we want to construct a second order ordinary dif\/ferential equation in $M$ we have to give a section of $E(TM)$. From now on we shall indicate an element of $E(TM)$ by $(v,v,a)$ using the lower case letter ``$a$" to signify acceleration. One has a canonical map
$\tilde \pi: T^*(TM)\to T^*M$, given in local coordinates by
\[
\tilde \pi:(q,v; A,B) \mapsto (q,B).
\]

One has a subbundle  $Z^*(TM) \subset T^*(TM)$ consisting of
forms $A$ such that $\tilde \pi(A) = 0$. This subbundle allows for
another map $\pi^\flat:Z^*(TM)\to T^*M$ which in local coordinates is given by
\[
(q,v; A,0)\mapsto (q, A),
\]
which can be seen to be consistent with coordinate changes by another tedious and unenlightening exercise.

A {\em Lagrangian\/}, conventionally expressed as $L(t,q,\dot q)$ is to be though of as a time-dependent function on $TM$, that is $L:\lR\times TM\to \lR$. There is a formal device by which a time-dependent Lagrangian can be replaced by an equivalent time-independent Lagrangian in  $T\lR\times TM'$ with another manifold $M'$, which we explain later. For what we do now, it's more convenient to treat the time-independent case and so we assume, until further notice and  without loss of generality, that $L$ is time-independent.
Given $L$ there is now the  map $p_L:TM\to T^*M$ which in local coordinates is given by
\[
p_L(q,v) = \left(q,\frac{\partial L}{\partial v^j}\,dq^j\right)= \left( q,\pi(L)_i\,dq^i\right)
\]
but is a coordinate independent construct. This is the familiar ``conjugate momentum". Familiar but not quite innocent, its coordinate independent def\/inition is
\[
p_L=\pi\circ\Lambda\circ dL,
\]
where $\pi:T(T^*M)\to T^*M$ is the canonical projection. Thus without $\Lambda$ one would not have conjugate momenta, nor the much traveled bridge between Lagrangian and Hamiltonian systems.

Euler's equations whose coordinate version is conventionally written as
\[
\pd L{q^i}-\frac d{dt}\pd{L}{\dot q^i}=0,
\]
 in coordinate independent notation can be shown to express a condition
 on a vector $X=(v,v,a)$ over a point $q$ in $E(TM)$, namely
\[
\Lambda \circ dL-(p_L)_*(X)=0.
\]
Here $p_{L*}$ is the dif\/ferential of $p_L$, this is where the second derivatives of $L$ appear. Note again the role of $\Lambda$. Under a certain condition of regularity, obeyed by most systems without constraints, this condition def\/ines a unique section of $E(TM)$ and thus a system of second-order dif\/ferential equations on $M$.

Equivalently, Euler's equation can be seen as the condition $EL(X)=0$ where $EL$ is the bundle map $E(TM)\to T^*(M)$ given by
\[
EL(X) =  \pi^\flat (dL - \Lambda^{-1}(p_{L*}X)),
\]
where an easy exercise in local coordinates shows that $dL - \Lambda^{-1}(p_{L*}X)\in Z^*(TM)$.

We can thus identify $EL$ as an element of a bundle of bundle morphism
\[
EL\in {\rm Hom}(E(TM),T^*M).
\]

In local coordinates the one-form $EL(X)$ is given by
\[
EL(X)=EL_i(X)\,dq^i=\left(\frac{\partial L}{\partial q^k}-\frac{\partial^2L}{\partial q^i\partial v^k}v^i-
\frac{\partial^2L}{\partial v^j\partial v^k}a^j\right)\,dq^k.
\]

We see from this that $EL$ in a coordinate basis is built up from the f\/irst and second partial derivatives (in the same coordinate basis) of $L$. These derivatives constitute coordinates of the second jet of $L$ and so the {\em construction\/} of the Euler equations is a bundle map
\begin{gather}\label{emap}
\cE:J^2(TM)\to {\rm Hom}(E(TM),T^*M).
\end{gather}
Concerning this map, there are two things to point out: (1) The bundles on both sides of (\ref{emap}) are {\em natural bundles\/}~\cite{kola-etal}, that is, dif\/feomorphism of the base manifold $M$ lift to bundle maps (which are also dif\/feomorphisms), and (2) The map $\cE$ is {\em equivariant\/}, that is, $ \Phi^\sharp \circ \cE = \cE \circ  \Phi^\sharp $ where $\Phi $ is a dif\/feomorphism of $M$ and $\Phi^\sharp $ is its lifting to ${\rm Hom}(E(TM),T^*M)$ on the left-hand side of the equation and to $J^2(TM)$  on the right-hand side.

One can now ask the natural question: what are all the equivariant maps between the two natural bundles that appear in~(\ref{emap})? We show below that these are very few, def\/ined by three constructs of one-forms, among which are the canonical momentum and the  Euler construct. We will supply a ``physicist's proof" of this result, meaning  a discussion about the possible ways of contracting indices among quantities that transform under the dif\/feomorphism group. A true mathematical proof using invariance theory is still being worked out and so we state our results as conjectures, though the ``physicist's proof" is generally a reliable method of quickly reaching the same result, providing thus strong evidence for the truth of the conjectures.

\section{Equivariance}\label{sec:equiv}

We work in a local coordinate system. The second jet of $L$ has the following coordinates induced via the local coordinates we are using:
\begin{enumerate}\itemsep=0pt
\item[1)] the function $L$;
\item[2)]  the f\/irst partial derivatives of $L$,
\[\pd L{q^i},\qquad \pd L{v^i};\]
\item[3)] the second partial derivatives of $L$
 \[\pd{{}^2 L}{ q^i\partial q^j},\qquad\pd{{}^2 L}{ q^i\partial v^j},\qquad\pd{{}^2 L}{ v^i\partial v^j}.\]
\end{enumerate}
Also $X=(v,v,a)\in E(TM)$ has components:
\begin{enumerate}\itemsep=0pt
\item[1)] the velocities $v^i$ which in Euler's dif\/ferential equation become $  \frac d{dt}q^i$;
\item[2)] the accelerations $a^i$ which in Euler's dif\/ferential equation become $  \frac d{dt}v^i=\frac{d^2}{dt^2}q^i$.
\end{enumerate}
These have various transformation properties in relation to a dif\/feomorphism of $M$, which locally we can take as a change of coordinates $q^i\to \tilde q^i$. Obviously $L$ is a scalar and $v$  is a vector. The  transformation law of the other quantities in $j^2L$ can be obtained from the relation
\[\tilde L(\tilde q^i, \tilde v^j)=L\left(q^i, \pd{q^j}{\tilde q^b}\tilde v^b \right),\]
where on the right-hand side $q$ is considered a function of $\tilde q$. We f\/ind
\begin{gather}\label{tldtq}
\pd{\tilde L}{\tilde q^i} = \pd{L}{q^k}\pd{q^k}{\tilde q^i}+\pd{L}{v^j}\ppdd{q^j}{\tilde q^i}{\tilde q^k}\tilde v^k, \\ \label{tldtv}
\pd{\tilde L}{\tilde v^i} = \pd{L}{v^k}\pd{q^k}{\tilde q^i},\\ \nonumber
\ppdd{\tilde L}{\tilde q^i}{\tilde q^j} =  \pd{L}{q^a}\ppdd{q^a}{\tilde q^i}{\tilde q^j} + \pd{L}{v^a} \pppddd{q^a}{\tilde q^i}{\tilde q^j}{\tilde q^b}{\tilde v}^b   + \ppdd{L}{q^a}{q^b}\pd{q^a}{\tilde q^i}\pd{q^b}{\tilde q^j} \\ \nonumber
\phantom{\ppdd{\tilde L}{\tilde q^i}{\tilde q^j}=}{} + \ppdd{L}{q^a}{v^b}\left[\pd{q^a}{\tilde q^i}\ppdd{q^b}{\tilde q^j}{\tilde q^c}+ \pd{q^a}{\tilde q^j}\ppdd{q^b}{\tilde q^i}{\tilde q^c}\right]{\tilde v}^c  + \ppdd{L}{v^a}{v^b}\ppdd{q^a}{\tilde q^i}{\tilde q^c}\ppdd{q^b}{\tilde q^j}{\tilde q^d}{\tilde v}^c {\tilde v}^d, \\
\label{tlddtqtv}
\ppdd{\tilde L}{\tilde q^i}{\tilde v^j} =  \pd{L}{v^a}\ppdd{q^a}{\tilde q^i}{\tilde q^j}+ \ppdd{L}{q^a}{v^b} \pd{q^a}{\tilde q^i}\pd{q^b}{\tilde q^j}+\ppdd{L}{v^a}{v^b}\ppdd{q^a}{\tilde q^i}{\tilde q^c}{\tilde v}^c\pd{q^b}{\tilde q^j},\\ \label{tlddtvtv}
\ppdd{\tilde L}{\tilde v^i}{\tilde v^j} =  \ppdd{L}{v^a}{v^b}\pd{q^a}{\tilde q^i}\pd{q^b}{\tilde q^j}.
\end{gather}

The components of $X$ transform as
\begin{gather*}
\tilde v^i  =  \pd{\tilde q^i}{q^a}v^a, \\
\tilde a^i = \pd{\tilde q^i}{ q^j}a^j+\ppdd{\tilde q^i}{q^a}{q^b}v^av^b.
\end{gather*}

We shall also need an expression for the  second derivatives of $\tilde q^i$ with respect to $q^i$ in terms of the other order of coordinates
  \[
 \ppdd{\tilde q^i}{q^k}{q^l}=-\ppdd{q^a}{\tilde q^b}{\tilde q^c}\pd{\tilde q^a}{q^k} \pd{\tilde q^b}{ q^l} \pd{\tilde q^i}{q^a}.
 \]

So the problem now becomes: how do we put all the above ingredients together to get a~one-form?

Some of the transformation pattern above are quite familiar: $L$ is a scalar, $ \pd L{v^i}$ is a co-vector and $ \ppdd{L}{v^i}{v^j}$ is a symmetric contravariant tensor of rank $2$. The  one form $ \pd L{v^i}\,dq^i$ is, as was already mentioned, the canonical momentum. Since $v^i$ is a vector, one can form another scalar $ \pd L{v^i}\,v^i$  which along with $L$ constructs the Hamiltonian $  H=\pd L{v^i}\,v^i-L$ (def\/ined in $TM$). The symmetric tensor $ \ppdd{L}{v^i}{v^j}\,dq^i\otimes dq^j$, if not degenerate, assures the regularity of Euler's equations by providing a unique section of $E(TM)$. There is a third scalar given by $ \ppdd{L}{v^i}{v^j}v^iv^j$ which is not as widely used as the other three. A fourth scalar is given by $\langle EL(X),v\rangle=EL_i(X)v^i$, the contraction of the Euler one-form with $v$. We conjecture that all other scalars are functions of these four.

The above transformations express an action of the dif\/feomorphism group which we now take to be on $J^2(TM)\times E(TM)$ and we are asking for an equivariant bundle map $J^2(TM)\times E(TM)\to T^*(M)$. This is equivalent to our previous request.
Let $\Omega =\omega _i\,dq^i$ be the putative one-form constructed in an equivariant way from the above data.
 Now because $  \ppdd L{\tilde q^i}{\tilde q^j}$ involves the third derivatives of $q^i$ and no other term does, $\Omega $ cannot depend on this element of $j^2L$. None of the other components in the $\tilde q^i$  coordinates receive contribution from $  \ppdd L{q^i}{q^j}$ so they form an invariant set of components and it is consistent to assume $\Omega $ is built only from this set. This set is still reducible as both the $ \pd L{v^i}$ and the $ \ppdd{L}{v^i}{v^j}$ form  invariant sets. The f\/irst of these gives rise to the canonical momentum one-form and the second to the one-form  $  \eta(L)= \ppdd{L}{v^i}{v^j}v^i\,dq^j$.  If $\Omega $ is not to be a combination of the canonical momentum and $\eta(L)$, then it must depend on either $  \pd{ L}{ q^i}$ or $  \ppdd{L}{q^i}{v^j} $ or both.

The second derivatives of the $q^i$ in the transformations of these components can only be compensated by a contraction of some of these with $\tilde a^i$. To facilitate this analysis choose a~dif\/feomorphism which f\/ixes a point $q_0$ in $M$ and at which  $  \pd{q^i}{\tilde q^j}$ is the identity matrix. By Borel's lemma $  A^i_{jk}=\ppdd{q^i}{\tilde q^j}{\tilde q^k}$ is an arbitrary set of components provided it is symmetric under interchange of $j$ and $k$. We have $\tilde v^i=v^i$,  $ \tilde A^i_{jk}=\ppdd{\tilde q^i}{ q^j}{ q^k}=-A^i_{jk}$, and $\tilde a^i=a^i-A^i_{ab}v^av^b$.

With this the transformation equations (\ref{tldtq})--(\ref{tlddtvtv}) become
\begin{gather*}
\pd{\tilde L}{\tilde q^i} = \pd{L}{q^i}+\pd{L}{v^j}A^j_{ik} v^k, \\
\pd{\tilde L}{\tilde v^i} = \pd{L}{v^i},\\
\ppdd{\tilde L}{\tilde q^i}{\tilde v^j} =  \pd{L}{v^a}A^a_{ij}+ \ppdd{L}{q^i}{v^j}+\ppdd{L}{v^a}{v^j}A^a_{ic}{\tilde v}^c,\\
\ppdd{\tilde L}{\tilde v^i}{\tilde v^j} =  \ppdd{L}{v^i}{v^j}.
\end{gather*}

Now the f\/irst two transformed jet elements above have only one free index and can suf\/fer no contraction while the other two  can each suf\/fer a contraction either with $\tilde v^i$ or $\tilde a^i$. It is now an easy exercise that the only combination of all these possible terms in which $A$ cancels out and which contains no terms proportional to $\pi(L)$ or $\eta(L)$ is precisely, up to a multiple, the Euler one-form which we know is an equivariant construct. From what was shown above we can now state:

\medskip

\noindent
{\bf Conjecture.} {\it The equivariant bundle maps in \eqref{emap} are of the form
\begin{gather}\label{equavs}
\cE(J^2L)(X)=\alpha \pi (L)(X)+\beta \eta(L)(X) + \gamma EL(X),
\end{gather}
 where $\alpha $, $\beta$ and $\gamma$ are functions of the four scalars mentioned above.
In this expression the first two terms depend only on the component $v$ of $X$.}

\section{Noether's theorem}\label{sec:noeth}

We consider all Lagrangian theories as ef\/fective theories arising from a fundamental theory by relegating degrees of freedom to the background. The set of ef\/fective theories form something like a partially ordered set by which one theory is related to another if the former is an ef\/fective version of the latter. In the quantum version, Feynman's integral provides a mechanism for forming ef\/fective theories by integrating over the degrees of freedom one wishes to suppress and  rewriting the rest in term of those one wishes to promote (we discuss this in Section~\ref{sec:conc}). With luck one passes from one Lagrangian theory to another with fewer degrees of freedom. The inverse process of ``de-ef\/fectivation'' of a theory has not received mathematical attention though it has historic precedence. Passing from the Fermi theory of weak interactions to the Weinberg--Salam electroweak theory is a prime example. The common practice of introducing new degrees of freedom to either simplify the treatment or make a given Lagrangian theory conform better to one's designs calls attention to the importance of this process.
A good mathematical treatment of ``de-ef\/fectivation'' is long overdue.

In our search for a replacement for the variational calculus we shall take the attitude that if the  introduction of  new degrees of freedom in such ``de-ef\/fectivations'' leads to a simplif\/ied perspective, then this perspective should be the one adopted. We justify this by noting that the suppression of degrees of freedom can lead to a theory in which certain simple relations in the original can assume less transparent form in the new.  One is thus not trying to replace {\em  all\/} Lagrangian theories and the concomitant variational calculi with something else, only those that exhibit certain simplicity in relation to those that arise from them by passing to ef\/fective or equivalent versions, with fewer degrees of freedom. This will become clearer with explicit examples below.

We admit temporarily that $L$ could depend on $t$. A variation is conventionally written as
 \[t\mapsto t+\delta t=t+\omega \tau,\qquad q^i\mapsto q^i+\delta q^i=q^i+\omega \eta^i\]
with { $\omega $} inf\/initesimal and $\tau $ and $\eta ^i$ functions on $\lR\times M$, depending on time and position, but not on velocities. This lifts to a vector f\/ield on { $\lR \times T(M)$} (see Olver \cite{olv, Olver} for this and other constructs we do below)
\begin{gather}\label{varvecf}
\cX=\tau D+\epsilon^i\frac{\partial}{\partial
q^i}+(D\epsilon^i)\frac{\partial}{\partial v^i},
\end{gather}
 where  { $\epsilon^i=\eta^i-v^i\tau$} and { $D$} is the
total derivative
\[D=\pd{}t+v^i\pd{}{q^i}+a^i\pd{}{v^i}.\]
The vector f\/ield $\cX$ decomposes conveniently as $\cX_H+\cX_V$ where $\cX_H=\tau  D$ is known as the {\em horizontal\/} component and the rest as the {\em vertical\/} component.
Note that $a^i$ refers to the acceleration as a component of an element of $E(TM)$. This means that $D$ is not a f\/ield on $\lR\times TM$ and so at f\/irst sight neither would be $\cX$, but the  $a^i$ contributions from the two terms containing~$D$ in~(\ref{varvecf}) cancel out. The use of $D$ simplif\/ies many expressions and is a convenient device.

Let now $S=\int L(t,q(t),\dot q(t))\,dt$ be the action integral. The variation of $S$ is then
\[
\delta S= \int \pounds_\cX(L\,dt)=\int (\cX(L)dt+ L\pounds_\cX(dt)),
\]
where $\pounds_\cX$ is the Lie derivative with respect to $\cX$.

One f\/inds after a short calculation that
\begin{gather}\label{lieldt}
\pounds_\cX(L\,dt)=\left(\epsilon^i EL_i(X)+D\left(\epsilon^i\frac{\partial L}{\partial v^i} +L\tau\right)\right)\,dt +L\frac{\partial \tau}{\partial q^i}(dq^i-v^i\,dt),
\end{gather}
where the f\/irst term is $\cX(L)dt$ and the second $ L\pounds_\cX(dt)$. In spite of the presence of the acce\-le\-ra\-tion $a^i$ in the element $X$ of $E(TM)$ and in $D$, these contributions cancel out from the full expression, though present in the individual contributions, an important fact. Concerning the coef\/f\/icient of $dt$ in the f\/irst term we have
\begin{gather}\label{nomore}
\epsilon^i EL_i(X)+D\left(\epsilon^i\frac{\partial L}{\partial v^i} +L\tau\right)=\cX_V(L)+D(L\tau),
\end{gather}
a fact that we shall use below.
The equality of the two expressions makes use of the fundamental isomorphism $\Lambda $.

The usual statement of the Noether  theorem is that if $\delta S=0$ integrated over an arbitrary interval, then the solutions of Euler's equations satisfy a conservation law. Translated into our language this means that if $\pounds_\cX(L\,dt)=0$ then solutions satisfy a conservation law. Indeed in a pull-back  of (\ref{lieldt}) onto an integral curve $(q(t),v(t),v(t),a(t))\in E(TM)$ of Euler's equations, the second term vanishes since $  v^i(t)=\frac{dq^i(t)}{dt}$ and the pullback of $\epsilon ^iEL_i(X)\,dt$ vanishes because $EL_i(X)=0$ is precisely Euler's equations. Thus we have on such integral curves that $  D\left(\epsilon^i\frac{\partial L}{\partial v^i} +L\tau\right)=0$ which is a conservation law.

If one is to replace variational calculus by a purely geometric formalism one would expect to state Noether symmetries (the analog of $\delta S=0$) purely by  $\cX(L)=0$ and deduce conservation laws from this. From (\ref{lieldt}) we see that this would be the case if  { $\tau=0$}. Now from the point of view of an underlying fundamental theory, the introduction of $\delta t$ along with $\delta q^i$ is seemingly contradictory. Under the relational view of space-time, the time and space coordinates are nothing more than constructs from events, which are governed by fundamental degrees of freedom. Varying these degrees of freedom would bring as a consequence a variation of the space-time coordinates and these should not have an independent variation. Thus one {\em should\/} relate our variational calculation above to one in which $\tau =0$ and treat the integration variable as a mere parameter.
We achieve this by a ``de-ef\/fectivation`` of $L$: Promote  $t$ to a dynamical variable (think of it as $q^0$) and let $s$ be the integration variable. Since $t$ depends on $s$ we need also introduce the ``velocity of time", that is $  w=\frac{dt}{ds}$.  Let $  \hat v^i=\frac{dq^i}{ds}$. One has, going back to the integral
\[
\int L(t,q(t),v(t))\,dt =\int L\left(t(s), q(s),\frac{\hat v(s)}{w(s)}\right)w(s)\,ds.
\]
One should now in principle consider the Lagrangian function $L(t, q,{\hat v}/{w})w$. This isn't quite right as now $t$ is to be an arbitrary function of~$s$ making it a~gauge variable but we don't have a~gauge theory (variations with respect to~$t$ will impose restriction we don't want). The way out is the oft used trick of gauge f\/ixing. Introduce yet another dynamic variable $\lambda$ (think of it as~$q^\infty$) as a Lagrange multiplier to f\/ix the gauge to $w=1$ and thus use the Lagrangian
\[
\hat L(t,q,\lambda, \hat v)= L(t, q,{\hat v}/{w})w + \lambda(w-1).
\]

\def\hati#1{#1(t,q,\hat v/w)}
Everything works out perfectly now. The variation with respect to the $q^i$ variables gives
\[
E_i(\hat L)=\hati{w\pd L{q^i}}-\frac d{ds}\left(\hati{\pd L{v^i}}\right).
\]
If we divide the right-hand side by $w$ and equate the result to zero we get the re-parameterized (with $s$ as independent variable) version of the original Euler equations.

The variation  with respect to $q^0=t$ gives
\[
E_0(\hat L)=\hati{w\pd L{t}}-\frac d{ds}\left(\hati L-\frac1w\hati{\pd L{v^i}}v^i+\lambda\right),
\]
whose vanishing def\/ines $\lambda$ up to a constant as a function of the other variables
\[\frac d{ds}\lambda =\hati{w\pd L{t}}+\frac d{ds}\left(\frac1w\hati{\pd L{v^i}}v^i-\hati L\right).\]
Now $\lambda$ has no conjugate momentum and this theory is thus one with constraints (in the Dirac sense).

The variation with respect to $q^\infty=\lambda$ gives
\[
E_\infty(\hat L)= w-1,
\]
whose vanishing f\/ixes the gauge and forces $w=1$, or in other words $t=s+t_0$, with $t_0$ another constant of integration.

This theory therefore is equivalent to the original one modulo the trivial freedom of choosing the integration constants for $\lambda$ and $t$.

One has to now check if the variational symmetries of the two theories are equivalent. In terms of the original $\tau(t,q)$ and $\eta(t,q)$ we now have a new vector f\/ield (no further terms will be necessary)
\begin{gather}\label{newx}
\hat\cX =  \tau\pd{}t+\eta^i\pd{}{q^i}+\eta^\infty\pd{}{\lambda}+\hat D\tau\pd{}w+\hat D\eta^i \pd{}{\hat v^i},
\end{gather}
where we have the new total derivative
\[\hat D=\pd{}s+w\pd{}t+\hat v^i\pd{}{q^i}\]
(no further terms will be necessary) and  where $\eta^\infty$ is the yet to be discovered variation of $\lambda$ ($\delta\lambda=\omega\eta^\infty$, $\omega$ the inf\/initesimal). Note there is no $\partial/\partial s$ term in (\ref{newx}) meaning that $\hat \tau$, the {\em  new\/} $\tau$ is zero. Likewise the new $\epsilon^i$ functions given by $\hat\epsilon^i=\hat\eta^i-\hat v^i\hat\tau$ coincide with the $\hat\eta^i=\eta^i$ functions; also $\hat\epsilon^0=\eta^0=\tau$ and $\hat\epsilon^\infty=\hat \eta^\infty=\eta^\infty$. We now argue for the new action integral  $\hat S=\int \hat L \,ds$ that $\delta \hat S=0\Leftrightarrow \hat\cX(\hat L)=0$ after a choice for $\eta^\infty$. Referring to (\ref{nomore}) one f\/inds
\begin{gather}\label{badbad}
\cX_V(L)+D(\tau L)=\tau\pd Lt+\eta^i\pd L{q^i}+(D\eta^i-v^iD\tau)\pd L{v^i}+(D\tau)L.
\end{gather}
Also
\begin{gather}\label{goodgood}
\hat\cX(\hat L)=w\tau\pd Lt+w\eta^i\pd L{q^i}+(\hat D\eta^i-\frac{\hat v^i}w\hat D\tau)\pd L{v^i}+(\hat D\tau)L +w\eta^\infty +(\hat D\tau)\lambda,
\end{gather}
where all the derivatives of $L$ are to be evaluated at $(t,q,\hat v/w)$.

Let now  $\sigma(t)$ denote  the triple $(t,q(t),v(t))$ and similarly $\hat \sigma(s)$ the triple $(t(s),q(s),\hat v(s)/w(s))$. If $F(t,q,v)$ is any function and $\tilde F(t,q,\hat v)=F(t,q,\hat v/w)$ then one has
\[(DF)(\sigma(t))=\frac d{dt}(F(\sigma(t)))=\frac1{w(s)}\frac d{ds}(\tilde F(\hat\sigma(s)))=\frac1{w(s)}(\hat D\tilde F)(\hat\sigma(s)),
\]
where the $t$ on one side and the  $s$ on the other are related by their functional dependence $t(s)$. Keeping track of what's a function of what and make appropriate use of the chain rule one can set $\hat D=wD$ and $\hat v/w=v$. Now in view of (\ref{badbad}),  we can write (\ref{goodgood}) as
\[\hat\cX(\hat L)=w(\cX_V(L)+D(\tau L)+\eta^\infty +(D\tau)\lambda)\]
and so the two theories have the same variational symmetries if we def\/ine the variation of the new variable $\lambda$ as
\[
\eta^\infty=-(D\tau)\lambda.
\]

A remark is in order about this. Expanding one has
\[\eta^\infty=-(\partial \tau/\partial t+v^i\partial \tau/\partial q^i) \lambda=-(\partial \tau/\partial t+(\hat v^i/w)\partial \tau/\partial q^i)\lambda\]
and so this variation in general {\em  depends on the velocities\/} (of $q$ and of $t$).  In general such a~situation leads to an inf\/inite regress needing to compute the variations of ever higher derivatives of the variables involved, placing the problem in an inf\/inite dimensional jet space. But,  in certain circumstance this may not be the case and here this doesn't happen  as there are no loops of dependencies (that is, the variation of A depending on the variable B whose variation depends on the variable C etc., leading back to~A) in which a dependence on a velocity appears in each step (the inf\/inite regress stems from such loops)~\cite{ottsvet}. Now $\eta^\infty $ depends on the derivative of the~$q^i$  and~$t$ but the $\eta^i$ and $\tau$ don't depend on $\lambda$ or any of its $s$-derivatives at all and so we are thus safe from the inf\/inite regress.

We have thus come to our f\/irst conclusion: under an appropriate ``de-ef\/fectivation'' a Lagrangian can be assumed to be time independent without sacrif\/icing Euler's equations or Noether symmetries. This justif\/ies our assumption  of time-independent Lagrangians in Sections~\ref{sec:geom} and~\ref{sec:equiv}.

Concerning Noether conservations laws there is another situation called ``quasi-invariance'' meaning that $\delta S$ is not zero but an integral of a total derivative $\delta S=\int D\Phi \,dt$. One still deduces a conservation law for solutions of Euler's equation which now is  $D(\epsilon^i(\partial L/\partial v^i)+L\tau-\Phi)=0$. We now show that this too can be subsumed under simple invariance $\cX(L)=0$ under an appropriate ``de-ef\/fectivation''. Taking into account the f\/irst part of this section the context now is
 a Lagrangian $L(q,v)$ that is time-independent and  variations $\delta q^i=\omega\eta^i$, $\omega$ inf\/initesimal, and  $\delta t=0$.  Suppose now that $\cX(L)=D\Phi.$ By our assumption, $\eta ^i$, can only depend on $q^i$ as the integration variable is just a parameter, hence  $\Phi$ is a function only of $q^i$. As before  introduce now a new dynamic variable $q^0=\xi$ with velocity $v^0=\nu$ and another dynamic variable $q^\infty=\lambda$ whose velocity we'll not need. Consider the Lagrangian
\[
\hat L(q,v,\xi,\nu,\lambda)=L(q,v)+\nu D\Phi + \lambda(\xi-1).
\]
Variations with respect to  $q^i$ gives ( since $E(D \Phi)=0$)
\[E_i(\hat L)=E_i(L),\]
and we recover the old Euler equations.
Variations with respect to  $q^0$ gives
\[E_0(\hat L)=\lambda-DD\Phi\]
the vanishing of which def\/ines $\lambda$ in terms of the original dynamical variables $\lambda=DD\Phi$.
Variations with respect to  $q^\infty$ gives
\[E_\infty(\hat L)=\xi-1\]
the vanishing of which f\/ixes the new variable $\xi$ to be the constant $1$.

Again, as far as the dynamics is concerned we can consider $\hat L$ as def\/ining an equivalent theory.
Concerning Noether's theorem, the new vector f\/ield def\/ining the new variation has to be of the form
\[\hat \cX = \eta\pd{}q+D\eta\pd{}v+\eta^0\pd{}\xi+ \hat D\eta^0\pd{}\nu+\eta^\infty\pd{}\lambda,\]
where $\eta^0$ and $\eta^\infty$ are two new variations to be determined: $\delta\xi=\omega\eta^0$ and $\delta\lambda=\omega\eta^\infty$, and $\hat D$ is the new total derivative taking into account the new variables. One f\/inds
\[\hat\cX(\hat L)=\cX(L)+\nu \cX(D\Phi)+\eta^0\lambda+\hat D\eta^0D\Phi+\eta^\infty \xi.\]
Now for the quasi-invariance of $L$ to be equivalent to true invariance of $\hat L$ one need have
\[D\Phi+\nu \cX(D\Phi)+\eta^0\lambda+\hat D\eta^0D\Phi+\eta^\infty \xi=0.\]
There are seemingly many ways to achieve this, but a simple one is to take $\eta^0=0$ and \[\eta^\infty=-\frac1\xi(D\Phi+\nu\cX(D\Phi)).\]
Again, this is a variation that depends on velocities, but again, there is no problem.

The second conclusion is that conservations laws coming from quasi-invariance can be realized as coming from true invariance after an appropriate ``de-ef\/fectivation'' of the Lagrangian.

\section{Conclusions and the Feynman integral}\label{sec:conc}

The strange ef\/fectiveness of Lagrangian theories and the formal use  of variational calculus suggests that one should try to achieve the same results without recourse to the action integral and its variation. In this respect we have shown:
\begin{enumerate}\itemsep=0pt
\item Euler's equations can stand on their own as they arise from an equivariance principle as stated in the conjecture (\ref{equavs}). True, there are two other possible terms and one would like to be able to identify  just the Euler one-form in some canonical manner. In a sense this is possible for if $\ppdd{L}{v^i}{v^j}\neq 0$ then dif\/ferentiating (\ref{equavs}) with respect to the acceleration $a$ we can f\/ind $\gamma$ and so just the Euler term. This is not exactly a canonical identif\/ication, but is already progress in the right direction.
\item Variational symmetries can stand on their own. A variational symmetry is invariance of the Lagrangian under an inf\/initesimal dif\/feomorphism of the manifold $M$ lifted canonically to $TM$, {\em  provided\/} the Lagrangians are of a special class. Any Lagrangian can be ``de-ef\/fectivated'' to one in such a class.
\end{enumerate}

The action integral can now be viewed as a convenient short-cut to arrive at some purely geometric results. Its existence as a true integral, that is, as a number obtained by integrating an integrable function, is now seen to be irrelevant to the use to which it is put.
What is still left up in the air is why should there be Lagrangians at all. It's all well and good that equivariance and invariance lead to the usual variational results, but why start with a Lagrangian anyway? It seems that the Feynman integral can of\/fer some insight.
 One has:
\begin{gather}\label{fey}
Z=\int e^{iS(\phi)}\cD\phi,
\end{gather}
where $\phi$  stands for a set of f\/ields and $S(\phi)$  is the action integral $\int{\cal L}(\phi, \partial_\mu\phi)\,d^4x$. It is instructive to see how ef\/fective theories arise in this context.
  To get an ef\/fective theory out of (\ref{fey}) for some independent quantities $\psi$ that depend on the $\phi$ one then chooses further independent quantities~$\tilde \phi$ so that one can view the transition $\phi\mapsto (\psi,\tilde \phi)$ as a ``coordinate change in $\phi$ space''. One then has $\cD\phi = |\det(J)|\cD\psi\,\cD\tilde\phi$ where $J$ is the ``Jacobian matrix of the inverse coordinate change''. The ef\/fective theory for the variables $\psi$ is then given by
\[
Z_{\rm ef\/f}=\int e^{iS_{\rm ef\/f}(\psi)}\cD\psi,
\]
where
\[
e^{iS_{\rm ef\/f}(\psi)}=\int e^{iS(\psi,\tilde \phi)}|\det(J)|\cD\tilde\phi
\]
def\/ines the new ef\/fective action. The Feynman integral is thus a neat machine for getting ef\/fective theories: just change variables and do a partial integration. If $S$ is given by an action integral and one is lucky then $S_{\rm ef\/f}$ will also be given by an action integral of the {\em  effective\/} Lagrangian. If Feynman integration is the essence of quantum mechanics, then quantum mechanics has the enviable property that it allows any set of variables that you may chose to obey an ef\/fective theory that is also quantum mechanical in principle, though it may not seem so.
This general situation also explains why  macroscopic ef\/fective quantities (such as the Landau phase in superconductivity) do exhibit quantum behavior when the conditions are right (as in biased Josephson junctions). No quantity truly looses its quantum character and will exhibit it under the right conditions.

The use of the Feynman integral to create ef\/fective theory seems to beg the question of the need for Lagrangians, isn't there then some  ``fundamental Lagrangian'' from which all other theories will then be ef\/fective theories. Some people do search for this fundamental Lagrangian (a string-theorist will probably even exhibit his favorite, and there are various sums (discrete integrals) over combinatorial objects proposed for quantum gravity). Why should there be such a fundamental Lagrangian? Furthermore, the Feynman integral seems to give importance to the action integral, after all, it is the exponential of such that one is called to integrate.

Now nobody has ever succeeded in def\/ining the Feynman integral as a true integral in the measure-theoretic sense. If we start questioning the action integral as a fundamental ingredient in physical theories and begin to consider it as a mere expedient tool for expressing geometric relations, then one can question whether Feynman's integral is really about summing phases to calculate transition probabilities. Maybe it also is a short-cut expression for a construct that can be def\/ined otherwise. In  separate papers \cite{svetfmubs,svetwhy} we present exactly such an idea, that Feynman's integral is about the existence of mutually unbiased bases somehow related to causality. In a~f\/inite dimensional Hilbert space two bases $e_a$ and $f_b$ with $a,\,b=1,\dots,N$ are called {\em mutually unbiased\/} \cite{beng} if $  |(e_a,f_b)|^2=\frac 1N$. This means that knowing the result of a measurement in one of the bases gives no information about what the result of a subsequent measurement in the next basis. One then has
\begin{gather}\label{mub}
(e_a,f_b)=\frac{e^{iL(a,b)}}{\sqrt{N}}.
\end{gather}

Here one sees the appearance of the ``Lagrangian'' $L(a,b)$. It is constrained by the requirement that (\ref{mub}) be a unitary matrix. Seeing that the Feynman integral is an integral of phases, it can be viewed as the requirement of the existence of a certain system of mutually unbiased bases (interpreted appropriately in inf\/inite-dimensional Hilbert spaces), or approximates of such. Lagrangians are then the phase information carried in the inner product of eigenvectors taken from the two bases. This would explain the physical origin of Lagrangians and appropriate geometric principles would take care of variational calculus results.

Certain aspects of the choice of Lagrangians as practiced by physicists get suggestive cla\-ri\-f\/i\-ca\-tions from the idea that they are phases arising from inner products of mutually unbiased bases, or ones nearly so. This is especially true if one considers bases such as position or f\/ield-strength at two times with very small separation. From the positions (for f\/ield-strength analogous observations apply) at two times, in the limit of zero separation, one can construct a~position and a velocity and so the phase (Lagrangian) in this limit would be a function of position and velocity. This suggests why phase space is important and why f\/irst-order lagrangians seem to be of particular worth (one cannot deduce acceleration from two positions and a time dif\/ference). Another common requirement is that the Lagrangian (or better yet, the action integral) ought to be invariant under whatever symmetry group one feels governs the physics, or its description (as in gauge theories). This of course is obviously natural and need not be justif\/ied, however, in thinking of Lagrangian theories as being ef\/fective ones of some fundamental underlying one, and adopting the relational viewpoint of space-time, one is naturally led to dif\/feomorphism invariance (or better yet, equivariance) as a fundamental principle. It is in this scenario that the purely geometric ``variational calculus'' in integral-free terms should have its expression. If one can achieve this, one would surely be able to answer some of the questions posed at the beginning of this paper.

\subsection*{Acknowledgements}
This research was partially supported by the Conselho Nacional de Desenvolvimento Cient\'{\i}f\/ico e Tecnol\'ogico (CNPq), and the Funda\c{c}\~ao de Amparo \`a Pesquisa do Estado do Rio de Janeiro  (FAPERJ).

\pdfbookmark[1]{References}{ref}
\LastPageEnding


\begin{thebibliography}{99}

\footnotesize\itemsep=0pt


\bibitem{saun}Saunders~D.J., The geometry of jet bundles, Cambridge University Press, 1989.

\bibitem{olv}Olver P.J., Equivalence, invariants and symmetry, Cambridge University Press, 1995.

\bibitem{svetfmubs}Svetlichny~G., Feynman's integral is about mutually unbiased bases, \href{http://arxiv.org/abs/0708.3079}{arXiv:0708.3079}.

\bibitem{kola-etal}Kolar~I., Michor~P.W., Slovak~J., Natural operations in dif\/ferential geometry, Springer, New York, 1993, available at
\url{http://www.emis.de/monographs/KSM/}.

\bibitem{Olver} Olver P.J., Applications of Lie groups to
dif\/ferential equations,  Springer, New York, 1986.

\bibitem{ottsvet}Otterson~P., Svetlichny~G.,  On derivative-dependent inf\/initesimal deformations of dif\/ferentiable maps, {\it J.~Differential Equations} {\bf 36} (1980), 270--294.

\bibitem{svetwhy}Svetlichny~G., Why Lagrangians?, in Proceedings XXVI Workshop on Geometrical Methods in Physics
(July 1--7,  2007, Bialowieza, Poland), {\it AIP Conference Proceedings}, Vol.~956,
Editors P.~Kielanowski, A.~Odzije\-wicz, M.~Schlichenmeier and T.~Voronov, AIP, New York, 2007, 120--125.

\bibitem{beng} Bengtsson I., Three ways to look at mutually unbiased bases, \href{http://arxiv.org/abs/quant-ph/0610216}{quant-ph/0610216}.

\end{thebibliography}
\end{document}